\documentclass[9pt]{article}

\usepackage{cmumath}

\usepackage{thmtools}
\usepackage{thm-restate}
\allowdisplaybreaks

\usepackage{tikz}

\DeclareMathOperator{\Inf}{\bf Inf}
\DeclareMathOperator{\Stab}{\bf Stab}

\newtheorem{mydef}{Definition}[section]
\newtheorem{mytheorem}{Theorem}
\newtheorem{mylemma}[mytheorem]{Lemma}

\theoremstyle{remark}

\begin{document}
\begin{center}
{\Large A Noisy-Influence Regularity Lemma for Boolean Functions}\\
\begin{tabular}{c}
Chris Jones
\end{tabular}
\end{center}

\begin{abstract}

We present a regularity lemma for Boolean functions $f:\{-1,1\}^n \to \{-1,1\}$ based on noisy influence, a measure of how locally correlated $f$ is with each input bit. We provide an application of the regularity lemma to weaken the conditions on the Majority is Stablest Theorem. We also prove a ``homogenized" version stating that there is a set of input bits so that most restrictions of $f$ on those bits have small noisy influences. These results were sketched out by \cite{NoisyRL}, but never published. With their permission, we present the full details here.

\end{abstract}

\section{Introduction}
A recent theme in discrete mathematics has been the development of regularity lemmas, tools which break down large-scale combinatorial objects into a constant number of easy-to-understand pieces. The goal of this note is to prove such a regularity lemma for Boolean functions $f:\{-1,1\}^n \to \R$. 

There are two items to address. First, how does one ``break down" a Boolean function? We use a decision tree which queries individual bits at internal nodes and places a subfunction at each leaf. Furthermore, the depth of this decision tree will be independent of $n$. Second, what does it mean for a subfunction to be ``easy-to-understand"? One notion of ``easy-to-understand" that arises in other contexts is ``pseudorandom": possessing structure that is likely to arise if the object is chosen randomly. It is likely that a randomly-chosen Boolean function will not locally behave like any single input bit, and this is the notion of ``easy-to-understand" we adopt here. We will precisely define these concepts in Section~\ref{NoisyProof}, where we prove the main statement of the regularity lemma:

\begin{restatable}{mytheorem}{noisyreg}
\label{NoisyReg}
For every $\delta, \gamma  \in (0,1], \epsilon > 0$ and $f:\{-1,1\}^n \to \R$ such that $\E[f^2] \leq 1$, there is a decision tree~$\mathcal{D}$ of depth at most $1/\epsilon\delta\gamma$ and functions $f_L : \{-1,1\}^n \to \{-1, 1\}$ indexed by leaves $L$ of $\mathcal{D}$ such that 
\begin{enumerate}[(i)]
\item $f(x) = f_{\mathcal{D}(x)}(x)$

\item All but at most a $\gamma$ fraction of the $f_L$ have $(\epsilon, \delta)$-small noisy influences.

\end{enumerate}
\end{restatable}

We also prove a ``homogenized" version of the theorem. In this version, the decision tree must query the same bit on every level. The subfunctions of this tree correspond to the restrictions of $f$ on the queried inputs. Thus the theorem states that there are a constant number of bits so that most restrictions of $f$ have small influences:

\begin{restatable}{mytheorem}{noisyreghom}
\label{NoisyRegHom}
For every $\delta, \gamma  \in (0,1], \epsilon > 0$, $1/\epsilon\delta\gamma > 1$, and $f:\{-1,1\}^n \to \R$ such that $\E[f^2] \leq 1$, there is $J \subseteq [n]$ of size at most $2\uparrow\uparrow 1/\epsilon\delta\gamma$ such that all but at most a $\gamma$ fraction of restrictions of $f$ on $J$ have $(\epsilon, \delta)$-small noisy influences.
\end{restatable}

The class of low-influence functions is one way of representing real-world functions where each input contributes a small piece of the output. On the mathematical side the study of low noisy-influence functions has been driven by the development of invariance principles and connections between functions on the product probability space $\{-1,1\}^n$ and Gaussian space $\mathcal{N}(0,1)^m$ \cite{MOO:focs}. As an application of the regularity lemma to this study, we slightly weaken the conditions on the Majority is Stablest Theorem proven in \cite{MOO:focs}. The proof of Theorem~\ref{NoisyReg} and the application to the Majority is Stablest Theorem are based on sketches from \cite{NoisyRL}. The complete proofs here are presented with the permission of the authors.

\subsection{Previous Regularity Lemmas}

Regularity lemmas and decomposition results among different classes of Boolean functions are not new \cite{PTFRL}, \cite{AbelianRL}, \cite{LinRL}. Work of Ben Green from 2004 \cite{AbelianRL} established a regularity lemma for general abelian groups, which specializes to the Boolean case:
\begin{mytheorem}
\label{AbelianRL}
Let $f:\{-1,1\}^n \to \R$ such that $\E[f^2] \leq 1$, and let $\gamma \in (0,1], \epsilon > 0$. Then there is a ``generalized decision tree" $\mathcal{D}$ of height at most $1/\gamma\epsilon^2$ and a subfunction for each leaf $f_L:\{-1,1\}^n \to \R$ such that 
\begin{enumerate}[(i)]
\item $f(x) = f_{\mathcal{D}(x)}(x)$

\item All but at most a $\gamma$ fraction of leaves $L$ are $\epsilon$-regular: $\abs{\widehat {f_L}(S)}\leq \epsilon$ for every $S \neq \0$.

\end{enumerate}
\end{mytheorem}

A ``generalized decision tree" is permitted to split on parities of arbitrary subsets of bits, rather than parity of a single bit. In the original paper, Green furthermore obtained a generalized decision tree in which all nodes at the same level query the parity of the same set of bits. One can think of this as a sort of ``homogenized" version of the above theorem. 

One interpretation of such a ``homogenized" tree is that each leaf restricts the input to a different coset of a subspace (this subspace is the annihilator of all parity functions, with codimension the height of the tree), hence the following ``arithmetic regularity lemma":

\begin{mytheorem}
\label{ArithRL}
Let $f:\F_2^n \to \R$ such that $\E[f^2] \leq 1$, and let $\eps > 0$. Then there is a subspace $H$ of codimension at most $2\uparrow\uparrow 1/\gamma\eps^2$ such that at least $1-\gamma$ fraction of cosets $L$ of $H$ ensure that $f$ is $\eps$-regular on $L$.
\end{mytheorem}

Note the superexponential bound on the depth. Tower-type lower bounds have also been shown for Theorem~\ref{ArithRL} by Hosseini et al in \cite{AbelRLLower}.

A key philosophical difference between Green's work and ours is that special preference is given here to individual bits, and none to strings with Hamming weight greater than 1. In comparison, the arithmetic regularity lemma considers all nonzero elements as interchangeable vectors from $\mathbb{F}_2^n$. 

\section{Proof of Regularity Lemma}
\label{NoisyProof}
This section is concerned with proving Theorems~\ref{NoisyReg} and~\ref{NoisyRegHom}. First we establish some definitions. The terminology used is in alignment with \cite{AoBF}.
\subsection{Definitions}
Consider a Boolean function $f: \{-1,1\}^n \to \R$.

\begin{mydef}
\label{Stability}
The noise stability of $f$ at $\rho\in[0,1]$, written $\Stab_\rho[f]$, is $$\Stab_\rho[f] = \underset{\substack{(x,y)\\\rho- \text{correlated}}}{\E}[f(x)f(y)]$$ where $\rho$-correlated strings $(x,y)$ are formed by picking $x$ uniformly from $\{-1,1\}^n$, and $y$ by taking each bit $y_i$ to have correlation $\rho$ with $x_i$.
\end{mydef}

There is a Fourier formula for noise stability:
$$\Stab_\rho[f] = \displaystyle\sum_{S \subseteq [n]}\rho^{|S|}\widehat f(S)^2$$
From this we see, when $\rho \geq 0$, $\Stab_\rho[f] \geq 0$, and since $\rho \leq 1$, $\Stab_\rho[f] \leq \E[f^2]$.

\begin{mydef}
Define the $i^{th}$ directional derivative operator $D_i$ by 
$$(D_if)(x) = \frac{f(x^{(i \to 1)}) - f(x^{(i \to -1)})}{2}$$

\end{mydef}

\begin{mydef}
\label{NoisyInfluence}
For $\delta \in [0,1]$, the $(1-\delta)$-noisy influence of $x_i$ on $f$, written $\Inf^{(1-\delta)}_i[f]$, is $$\Inf^{(1-\delta)}_i[f] = \Stab_{1-\delta}[D_if]$$
\end{mydef}
$\Stab_{1-\delta}[f]$ is a measure of how locally constant $f$ is, and thus $\Inf_i^{(1-\delta)}[f]$ measures how much $f$ is locally correlated with the $i$-th input bit. Finally, low-influence functions are those that have all local correlations small:
\begin{mydef}
We say that $f$ has $(\epsilon, \delta)$-small noisy influences if $\Inf^{(1-\delta)}_i[f] \leq \epsilon$ for every $i$.
\end{mydef}

A decision tree is a particular representation of a Boolean function that computes input $x$ by querying for a particular input bit of $x$, and then proceeding to the left or right child depending on that value. In our decision trees, a leaf can contain a subfunction which will be evaluated on any inputs that evaluate to the leaf. For example, this decision tree has five subfunctions:

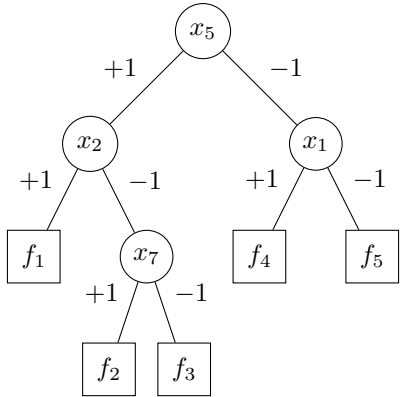
\begin{figure}[!ht]
  \centering
  \begin{tikzpicture}[
level 1/.style={sibling distance=30mm},
level 2/.style={sibling distance=15mm},
level 3/.style={sibling distance=10mm},
]
\node [circle,draw] (root){$x_5$}
  child {node [circle,draw] (1) {$x_2$}
    child {node [rectangle,draw, inner sep = 2pt, minimum width=7mm, minimum height = 7mm] (11) {$f_1$}
    }
    child {node [circle,draw, inner sep = 2pt, minimum width=7mm, minimum height = 7mm] (10) {$x_7$}
        child {node [rectangle,draw, inner sep = 2pt, minimum width=7mm, minimum height = 7mm] (101) {$f_2$}
        }
        child {node [rectangle,draw, inner sep = 2pt, minimum width=7mm, minimum height = 7mm] (100) {$f_3$}
        }
    }
  }
  child {node [circle,draw, inner sep = 2pt, minimum width=7mm, minimum height = 7mm] (0) {$x_1$}
    child {node [rectangle,draw, inner sep = 2pt, minimum width=7mm, minimum height = 7mm] (01) {$f_4$}
    }
  child {node [rectangle,draw, inner sep = 2pt, minimum width=7mm, minimum height = 7mm] (00) {$f_5$}
  }
  };
\path (root) -- (1) node [above left, midway] {$+1$};
\path (root) -- (0) node [above right, midway] {$-1$};
\path (1) -- (11) node [above left, midway] {$+1$};
\path (1) -- (10) node [above right, midway] {$-1$};
\path (0) -- (01) node [above left, midway] {$+1$};
\path (0) -- (00) node [above right, midway] {$-1$};
\path (10) -- (101) node [above left, midway] {$+1$};
\path (10) -- (100) node [above right, midway] {$-1$};
\end{tikzpicture}  
  \caption{An example decision tree.}
\end{figure}

We say that a decision tree is homogeneous if every level of the tree queries the same bit. The example is not homogeneous because both $x_2$ and $x_1$ are queried on level 1. Here and later we write $\mathcal{D}(x)$ for the leaf output by a decision tree.

\subsection{Proof of Theorems}
For clarity we restate the theorem to prove here.
\noisyreg*

When we say ``$\gamma$ fraction of leaves", we mean that making random decisions from the root of the tree leads to a leaf where the desired property holds with probability at least $1-\gamma$. 

The proof is constructive, and follows the energy increment technique used to prove other regularity lemmas. That is, if we have a decision tree $\mathcal{D}$ that computes $f$, we define the energy $\varphi(\mathcal{D})$ by
$$\varphi(\mathcal{D}) = \underset{L}{\E}[\Stab_{1-\delta}[f_L]]$$
We will show $0 \leq \varphi(\mathcal{D}) \leq 1$. If we at any time violate the goals of Theorem~\ref{NoisyReg} by having too many leaves with large noisy influences, we split each leaf on the bit with large influence to ``stabilize" those leaves. The proof strategy is to show that this increases the energy of $\mathcal{D}$ by a constant amount, and hence we won't have to repeat this splitting operation too many times.

The following equality will be used to show the energy change as we split on a bit with high influence, and hence is the crux of the argument:

\begin{mylemma} 
\label{EnergyInc}
Fix $i \in [n]$. Define $f_{-1}(x) := f(x^{(i \to -1)})$ and $f_1(x) := f(x^{(i \to 1)})$. Then
$$\frac{1}{2}\Stab_{1-\delta}[f_{-1}] + \frac{1}{2}\Stab_{1-\delta}[f_1] = \Stab_{1-\delta}[f] + \delta\Inf_i^{(1-\delta)}[f]$$ 
\end{mylemma}

{\it Proof.} 
The left hand side is $\underset{\substack{(x,y)\\(1-\delta)-\text{correlated}}}{\E}[f(x)f(y) \mid x_i = y_i]$. On the other hand, we can generate the same distribution by picking $(1-\delta)$-correlated strings $(x,y)$ and computing $X + Y$, where $X = f(x)f(y)$ and $Y$ is a correction factor that takes value $f(x)f(y^{(i \to x_i)}) - f(x)f(y)$. Note that $Y$ is only nonzero when $x_i \neq y_i$. Taking expectations and applying linearity, 
\begin{align*}
\E[X] &= \Stab_{1-\delta}[f]\\
\E[Y] &= \Pr[x_i \neq y_i]\left(\frac{1}{2}\E[Y\mid x_i = -1, x_i \neq y_i] + \frac{1}{2}\E[Y\mid x_i = 1, x_i \neq y_i]\right)\\
&= \frac{\delta}{2}\left(\frac{1}{2}\E[f(x^{(i \to -1)})f(y^{(i \to -1)}) - f(x^{(i \to -1)})f(y^{(i \to 1)})]\right.\\
&\qquad\left. + \frac{1}{2}\E[f(x^{(i \to 1)})f(y^{(i \to 1)}) - f(x^{(i \to 1)})f(y^{(i \to -1)})]\right)\\
& = \delta\E\left[\left(\frac{f(x^{(i \to 1)}) - f(x^{(i \to -1)})}{2}\right)\left(\frac{f(y^{(i \to 1)}) - f(y^{(i \to -1)})}{2}\right)\right]\\
&= \delta\E[D_if(x)D_if(y)] = \delta \Inf_i^{(1-\delta)}[f]
\end{align*}
\qed

Now we're ready to fill in the details of Theorem~\ref{NoisyReg}.

{\it Proof of Theorem~\ref{NoisyReg}.} We construct a decision tree $\mathcal{D}$ with the desired properties. Start $\mathcal{D}$ out as a \mbox{single-leaf} decision tree with $f$ itself at the leaf. 
We perform the following iterative splitting process on our decision tree $\mathcal{D}$: suppose we have a leaf $L$ and a bit $x_j$ such that $\Inf_j^{(1-\delta)}[f_L] > \epsilon$. Form decision tree $\mathcal{D}'$ by replacing $L$ with a query to $x_j$, and subfunctions $f_{L,-1}, f_{L, 1}$ defined by $f_{L,-1}(x) = f_L(x^{(j \to -1)})$ and $f_{L, 1}(x) = f_L(x^{(j \to 1)})$. How does the energy change? We claim it increases by at least $\epsilon\delta$ just on this leaf:
\begin{align*}
\varphi(\mathcal{D}') - \varphi(\mathcal{D}) &= \underset{L \sim \mathcal{D}'}{\E}[\Stab_{1-\delta}[f_L]] - \underset{L \sim \mathcal{D}}{\E}[\Stab_{1-\delta}[f_L]]\\
& = \Pr[\text{select leaf } L]\left(\frac{1}{2}\Stab_{1-\delta}[f_{L,-1}] + \frac{1}{2}\Stab_{1-\delta}[f_{L,1}] - \Stab_{1-\delta}[f_L]\right)
\end{align*}
By Lemma~\ref{EnergyInc} the latter quantity is
$$\Pr[\text{select leaf } L]\delta\Inf_j^{(1-\delta)}[f_L] > \Pr[\text{select leaf }L]\delta\epsilon$$
If there are at most $\gamma$ fraction of leaves that don't have $(\epsilon, \delta)$-small noisy influences, we are done. If not, performing the above replacement on each leaf, we replace $\mathcal{D}$ with our new decision tree $\mathcal{D}'$ such that $\varphi(\mathcal{D}') \geq \varphi(\mathcal{D}) + \epsilon\delta\gamma$ and $\mathcal{D}'$ has depth at most one greater than that of $\mathcal{D}$. Our next goal is to show termination by showing the energy is bounded.

Recall the definition of $\varphi(\mathcal{D})$, 
$$\varphi(\mathcal{D}) = \underset{L}{\E} [ \Stab_{1-\delta}[f_L]]$$
We have $\Stab_{1-\delta}[f] \geq 0$ (see Definition~\ref{Stability}), so averaging maintains $\varphi(\mathcal{D}) \geq 0$. We will show that $\varphi(\mathcal{D}) \leq 1$ to provide an upper bound on the energy.

Even if we hadn't chosen a bit with $\Inf_j^{(1-\delta)}[f_L] > \epsilon$, since $\Inf_j^{(1-\delta)}[f] \geq 0$, we still know that the energy does not decrease if we split on any leaf. During the iteration, a subfunction $f_L$ at depth $k$ fixes $k$ bits of $f$, no two subfunctions fix the same bits to the same values (they differ at their least common ancestor), and no root-to-leaf path splits on the same variable twice (once a variable $x_i$ has been split on, any subfunctions $f_L$ in that subtree are constant with respect to $x_i$, and hence have $\Inf_i^{(1-\delta)}[f_L] = 0$). From these three properties, we can extend our tree $\mathcal{D}$ via this splitting operation to a complete binary tree $\mathcal{T}$ of depth $n$, where each subfunction of $\mathcal{T}$ is constant and the subfunctions take on values $f(x)$ for each $x \in \{-1,1\}^n$. Nondecreasing energy upon splitting implies we can bound the energy of $\mathcal{D}$ by the energy of $\mathcal{T}$, which is
$$\varphi(\mathcal{T}) = \underset{L}{\E} [ \Stab_{1-\delta}[f_L]] = \underset{x\sim\{-1,1\}^n}{\E}[\Stab_{1-\delta}[f(x)]] = \underset{x\sim\{-1,1\}^n}{\E}[f(x)^2]\leq 1$$

Since $\varphi(\mathcal{D}) \leq 1$ at all times yet increases by $\epsilon\delta\gamma$, we iterate at most $1/\epsilon\delta\gamma$ times. This yields a bound on the depth of $\mathcal{D}$ of at most $1/\epsilon\delta\gamma$. The final decision tree $\mathcal{D}$ computes $f$, and has no leaf with large noisy influences. \qed

We can further enforce that the outcome be a homogeneous decision tree where every level of the tree queries the same bit. The relation between the previous theorem and the next is the same as the relation between Green's Theorem~\ref{AbelianRL} and Theorem~\ref{ArithRL}, and the proof adopts the same technique as in \cite{AbelianRL}.

\noisyreghom*

{\it Proof.} A homogeneous decision tree produced from Theorem~\ref{NoisyReg} will yield the desired $J$ by letting $J$ be all coordinates split on all levels. To produce such a decision tree we can perform the exact same iteration as in the proof of Theorem~\ref{NoisyReg}, with the following modification: suppose at level $k$ we have over $\gamma$ fraction of leaves that need to be split on high-influence variables $x_1, x_2, \dots, x_K$. Instead of splitting each leaf on one variable, we split every leaf on every variable from $x_1, x_2, \dots, x_K$, exactly in that order. Ignoring repeats, we may assume that the $x_1, x_2, \dots, x_K$ are all distinct. The energy difference due to a particular leaf $L$ is
$$\Pr[\text{select leaf }L]\left(\displaystyle\sum_y2^{-K}\Stab_{1-\delta}[f_{L, y}] - \Stab_{1-\delta}[f_L]\right)$$
where the $f_{L,y}$ range over all possible restrictions $y$ of the variables $x_1, x_2, \dots, x_K$. Notice that this is independent of the order in which the variables $x_1, x_2, \dots, x_K$ were split. Hence the energy difference is equal to that in which we first split $L$ on the $x_i$ with high influence, and then on the remaining $x_1, \dots, x_K$. This shows that the energy increase is again at least $\epsilon\delta\gamma$, though now every level queries the same variable.

Suppose the total depth of our decision tree after $k$ iterations is $d(k)$. We can have at most $K \leq 2^{d(k)}$ splits on the next iteration, hence the total depth of the tree satisfies
$$d(k+1) \leq d(k)  + 2^{d(k)}\qquad\qquad d(0) = 0$$
By induction we prove $d(k) \leq 2\uparrow\uparrow k -1$. Checking the induction step,
$$d(k+1) \leq  2\uparrow\uparrow k -1+\frac{1}{2}\cdot2\uparrow\uparrow (k+1) \leq \frac{1}{2}\cdot 2\uparrow\uparrow (k+1) -1+\frac{1}{2}\cdot2\uparrow\uparrow (k+1) $$ $$= 2\uparrow\uparrow (k+1) - 1 $$\qed

\section{Application: Quasirandom Functions and Majority is Stablest}
\label{Applications}

The (mean 0) Majority is Stablest Theorem, first proven in \cite{MOO:focs}, says that among Boolean functions with mean 0, Maj$_n$ has asymptotically the highest noise stability. In its full generality, the Majority is Stablest Theorem bounds the noise stability of an arbitrary Boolean function $f$ by a function dependent on $\E[f]$.

\begin{mydef}
Fix $\rho \in [0,1]$. The Gaussian quadrant probability $\Lambda_\rho(\mu): [0,1] \to [0,1]$ is defined by 
$$\Lambda_\rho(\mu) = \Pr[z_1 \leq t \cap z_2 \leq t]$$
where $z_1, z_2$ are standard Gaussians with correlation $\rho$, and $t$ is the inverse of the standard Gaussian CDF at $\mu$ i.e. $t$ is such that the area under the standard Gaussian and to the left of $t$ is $\mu$.
\end{mydef}

\begin{mytheorem} (General-Volume Majority Is Stablest Theorem) 
\label{MIST}
Let $f:\{-1,1\}^n \to [0,1]$ such that $f$ has $(\epsilon, \frac{1}{\log(1/\epsilon)})$-small noisy influences. Then for any $0 \leq \rho < 1$,
$$\Stab_\rho[f] \leq \Lambda_\rho(\E[f]) + O\left(\frac{\log \log(1/\epsilon)}{\log(1/\epsilon)}\right) \cdot \frac{1}{1-\rho}$$ 
\end{mytheorem}

Question: can we get by with a weaker notion of pseudorandomness than small noisy influences? We prove that the Majority is Stablest Theorem still holds if we replace ``small noisy influences" with ``small low-degree Fourier coefficients". 

\begin{mydef}
We say that $f: \{-1, 1\}^n \to \R$ is $(\epsilon, \delta)$-quasirandom if $\abs{\widehat f(S)} \leq \epsilon$ for $0 <|S| \leq 1/\delta$.
\end{mydef}

Informally, a quasirandom function is one which has small low-degree Fourier coefficients. It is strictly a weaker condition than having small (noisy) influences:

\begin{mylemma}
Suppose $f:\{-1,1\}^n \to \{-1,1\}$ has $(\epsilon^2, \delta)$-small noisy influences for some $\delta \in (0,1/2), \epsilon > 0$. Then $f$ is $(\epsilon, O(\delta))$-quasirandom.
\end{mylemma}

{\it Proof. } We go by contrapositive. Suppose there is a Fourier coefficient $\abs{\widehat f(S)} > \epsilon$ for some small $S, \abs{S} \leq 1/\delta$. For any $i \in S$, evaluating the Fourier formula for noisy influences on $D_if$,
$$\Inf^{(1-O(\delta))}_i[f] = \displaystyle\sum_{R \ni i} (1-O(\delta))^{|R|-1}\widehat f(R)^2$$for an appropriate linear factor $O(\delta)$ to be chosen later.  
$$\displaystyle\sum_{R \ni i} (1-O(\delta))^{|R|-1}\widehat f(R)^2\geq  (1-O(\delta))^{|S|-1}\widehat f(S)^2 > (1-O(\delta))^{1/\delta}\epsilon^2 \geq \epsilon^2$$
We choose the constant in $O(\delta)$ so that the last inequality holds. Divide all $\delta$ in the proof by the appropriate constant to prove the stated claim.
\qed

We will prove that the Majority is Stablest Theorem holds under the assumption that $f$ is $(o(1), o(1))$-quasirandom. That is, there are quasirandomness parameters that tend to 0 so that any $f$ satisfying those parameters also satisfy the Majority is Stablest inequality. Here is the generalization we prove:
\begin{mytheorem}
\label{QuasiMIST}
For $f:\{-1, 1\}^n \to [0,1]$ such that $f$ is $(\epsilon2^{-\frac{\log^2(1/\epsilon)}{\epsilon \log\log(1/\epsilon)}}, \epsilon\frac{\log\log(1/\epsilon)}{\log^2(1/\epsilon)})$-quasirandom, and for any~$0 \leq \rho < 1$,
$$\Stab_\rho[f] \leq \Lambda_\rho(\E[f]) + O\left(\frac{\log\log(1/\epsilon)}{\log(1/\epsilon)}\right) \cdot \frac{1}{1-\rho}$$
\end{mytheorem}

The following lemma is philosophically the reason for using the regularity lemma with quasirandom functions. Informally, quasirandom functions don't change their means too much on restriction. This is Proposition 6.12 of \cite{AoBF}.
\begin{mylemma}
\label{QuasiRestr}
Let $f: \{-1,1\}^n \to \R$ and $\epsilon \geq 0, \delta > 0$.
\begin{enumerate}[(1)]
\item If $f$ is $(\epsilon, \delta)$-quasirandom then any restriction of at most $1/\delta$ coordinates changes $f$'s mean by at most $2^{1/\delta}\epsilon$.
\item If $f$ is not $(\epsilon, \delta)$-quasirandom then some restriction to at most $1/\delta$ coordinates changes $f$'s mean by more than $\epsilon$.
\end{enumerate}
\end{mylemma}

\subsection{Proof of Theorem~\ref{QuasiMIST}}

{\it Proof.} Suppose $f:\{0,1\}^n \in [0,1]$ satisfies the conditions of Theorem~\ref{QuasiMIST}. By Theorem~\ref{NoisyReg} find a decision tree $\mathcal{D}$ computing $f$ of height at most $\frac{\log^2 (1/\epsilon)}{\epsilon\log\log(1/\epsilon)}$ so that, for all but at most a $\gamma = \frac{\log\log(1/\epsilon)}{\log(1/\epsilon)}$ fraction of leaves, subfunctions $f_L$ have $(\epsilon, \frac{1}{\log(1/\epsilon)})$-small noisy influences.

By the General-Volume Majority is Stablest Theorem (Theorem~\ref{MIST}), for all but at most $\gamma$ fraction of subfunctions $f_L$ we have 
$$\Stab_\rho[f_L] \leq \Lambda_\rho(\E[f_L]) + O\left(\frac{\log \log(1/\epsilon)}{\log(1/\epsilon)}\right) \cdot \frac{1}{1-\rho}$$

The functions $f_L$ are restrictions of $f$ to at most $\frac{\log^2 (1/\epsilon)}{\epsilon\log\log(1/\epsilon)}$ coordinates (the height of $\mathcal{D}$). By Lemma~\ref{QuasiRestr}, $\E[f_L]$ is at most 
$$2^{\frac{\log^2 (1/\epsilon)}{\epsilon\log\log(1/\epsilon)}}\left(\epsilon2^{-\frac{\log^2(1/\epsilon)}{\epsilon \log\log(1/\epsilon)}}\right) = \epsilon$$ away from $\E[f]$. Indeed, we could have picked any quasirandomness parameters in the statement of this theorem that ensured this bound.

$\Lambda_\rho$ is 2-Lipschitz (Exercise 11.19 of \cite{AoBF}), thus
$$\Stab_\rho[f_L] \leq \Lambda_\rho(\E[f]) + 2\abs{\E[f] - \E[f_L]} + O\left(\frac{\log \log(1/\epsilon)}{\log(1/\epsilon)}\right) \cdot \frac{1}{1-\rho} $$ $$\leq \Lambda_\rho(\E[f]) + O(\epsilon) + O\left(\frac{\log \log(1/\epsilon)}{\log(1/\epsilon)}\right) \cdot \frac{1}{1-\rho}$$

This bound is independent of $L$. By looking at the initial energy compared with the final energy in Theorem~\ref{NoisyReg},
$$\Stab_\rho[f] \leq \underset{L}{\E}[\Stab_\rho[f_L]]$$
For $\gamma$ fraction of leaves we can do no better than upper bound $\Stab_\rho[f_L] \leq \E[f_L^2] \leq 1$. For the rest of the leaves, we drop the $(1-\gamma)$ term and use our derived bound on $\Stab_\rho[f_L]$,
$$\Stab_\rho[f] \leq \gamma + \Lambda_\rho(\E[f]) + O(\epsilon) + O\left(\frac{\log \log(1/\epsilon)}{\log(1/\epsilon)}\right) \cdot \frac{1}{1-\rho}$$
$$ = \frac{\log\log(1/\epsilon)}{\log(1/\epsilon)} + \Lambda_\rho(\E[f]) + O(\epsilon) + O\left(\frac{\log \log(1/\epsilon)}{\log(1/\epsilon)}\right) \cdot \frac{1}{1-\rho}$$
$$\leq \frac{\log\log(1/\epsilon)}{\log(1/\epsilon)}\cdot\frac{1}{1-\rho} + \Lambda_\rho(\E[f]) + O(\epsilon)\cdot\frac{1}{1-\rho} + O\left(\frac{\log \log(1/\epsilon)}{\log(1/\epsilon)}\right) \cdot \frac{1}{1-\rho}$$
$$ = \Lambda_\rho(\E[f]) + O\left(\frac{\log \log(1/\epsilon)}{\log(1/\epsilon)}\right) \cdot \frac{1}{1-\rho}$$
where the last line follows because $\epsilon\to 0$ (indeed, any polynomial in $\epsilon$) much faster than $\frac{\log\log(1/\epsilon)}{\log(1/\epsilon)}$. \qed

\section{References}
\begingroup
\renewcommand{\section}[2]{}
\bibliographystyle{alpha}

\endgroup
\end{document}